\documentclass[graphicx]{iopart}
\usepackage{iopams,setstack}
\usepackage{epsfig}
\begin{document}

\title{Influence of asymmetry and nodal planes on high-harmonic generation in heteronuclear molecules}
\author{B. B. Augstein and C. Figueira de Morisson Faria}
\address{Department of Physics and Astronomy, University College
London, Gower Street, London WC1E 6BT, United Kingdom}
\date{\today}

\begin{abstract}
{The relation between high-harmonic spectra and the geometry of the
molecular orbitals in position and momentum space is investigated.
In particular we choose two isoelectronic pairs of homonuclear and
heteronuclear molecules, such that the highest occupied molecular
orbital of the former exhibit at least one nodal plane. The imprint
of such planes is a strong suppression in the harmonic spectra, for
particular alignment angles. We are able to identify two distinct
types of nodal planes. If the nodal planes are determined by the
atomic wavefunctions only, the angle for which the yield is
suppressed will remain the same for both types of molecules. In
contrast, if they are determined by the linear combination of atomic
orbitals at different centers in the molecule, there will be a shift
in the angle at which the suppression occurs for the heteronuclear
molecules, with regard to their homonuclear counterpart. This shows
that, in principle,  molecular imaging, which uses the homonuclear
molecule as a reference and enables one to observe the wavefunction
distortions in its heteronuclear counterpart, is possible. }
\end{abstract}
 \maketitle
\section{Introduction}
High harmonic generation, first discovered in the late 1980s
\cite{1987}, is a process which occurs when atoms or molecules are
exposed to intense laser fields, and results in the emission of high
frequency coherent radiation.  It can be readily understood in terms
of the three step model, \cite{Lewenstein,Corkum}, where an electron
is ionized from the atom or molecule, propagates in the laser field,
and recombines with its parent ion, upon which high harmonics are
released.

Currently, there is much interest in this process due to the
possibility of tomographic orbital reconstruction
\cite{Kanai,Itatani,Smirnova}, and quantum interference effects
\cite{doubleslit,Milosevic,Madsen}. In particular
quantum-interference minima and maxima due to high-harmonic emission
at spatially separated centers, and high-harmonic suppression due to
the presence of nodal planes in the orbital wavefunctions, have been
identified and discussed theoretically. Such studies have been
extensively performed in diatomic molecules
\cite{Smirnova,doubleslit,Milosevic,Madsen,tddft,CarlaBrad,BradCarla}.
In many of these studies so far, homonuclear molecules with definite
orbital symmetry, either gerade or ungerade, have mostly been used,
whereas heteronuclear molecules driven by strong fields are now
starting to attract attention \cite{Madsenhetero}.  The orbital
wavefunctions in such molecules do not have a definite orbital
symmetry, and heteronuculear molecules contain a static dipole
moment. These attributes will affect the harmonic spectra and other
strong field phenomena. Hence, a legitimate question to ask is
whether two-center interference patterns in the spectra, or the
imprint of nodal planes in form of high-harmonic suppression, can
still be observed in the spectra.

  In this work, we compare isoelectronic homonuclear and heteronuclear molecules, with particular emphasis on nodal planes in the position
  space wavefunction and two-center interference effects.  We assume the Born-Oppenheimer approximation to be valid,
  which should hold for large enough molecules \cite{Madsen}. Specifically, we take the pairs $\mathrm{Be}_2$ and LiB,
  and $\mathrm{O}_2$ and NF. The homonuclear molecule in former and latter pair have a $\sigma_{u}$ and $\pi_{g}$ highest
  occupied molecular orbital (HOMO), respectively. Such orbitals contain nodal planes which are altered when a heteronuclear
  counterpart is chosen. In our investigations, we employ the single-active electron and the strong-field approximation and assume
  that the HOMO is the only active orbital. The influence of multiple orbitals has been discussed elsewhere \cite{Smirnova,CarlaBrad,Patchkovskii}.

This work is organized as follows. In Sec. \ref{model} the three step model (Sec. \ref{saddlepoint}), molecular orbital construction (Sec. \ref{molecular}) and interference conditions (Sec. \ref{interf}) are discussed.  In Sec. \ref{results}, we present our results. This section is split into two main parts. First, in Sec. \ref{wfunctions}, we address the molecular orbitals obtained by us, with a particular emphasis on their nodal planes, or the absence thereof for heteronuclear molecules. Subsequently, in Sec. %
\ref{spectra} we discuss how such characteristics manifest themselves in the
spectra.  Finally, in Sec.\ref{conclusions} we state the main conclusions to be drawn from this work.

\section{Model}
\label{model}
\subsection{Saddle-point equations}
\label{saddlepoint}
{The transition amplitude for high harmonic generation, within the framework of of the strong field approximation (SFA) \cite{Lewenstein}, reads as,

\begin{eqnarray}
b_{\Omega } &\hspace{-0.1cm}=\hspace*{-0.1cm}&-i\int_{-\infty }^{\infty }%
\hspace*{-0.5cm}dt\int_{-\infty }^{t}~\hspace*{-0.5cm}dt^{\prime }\int
d^{3}ka_{\mathrm{rec}}^{\ast }(\mathbf{k}+\mathbf{A}(t))a_{\mathrm{ion}}(%
\mathbf{k}+\mathbf{A}(t^{\prime }))  \nonumber \\
&&\exp [iS(t,t^{\prime },\Omega ,\mathbf{k})]+c.c.,  \label{amplhhg}
\end{eqnarray}%
with the action
\begin{equation}
S(t,t^{\prime },\Omega ,\mathbf{k})=-\frac{1}{2}\int_{t^{\prime }}^{t}[%
\mathbf{k}+\mathbf{A}(\tau )]^{2}d\tau -E_{0}(t-t^{\prime })+\Omega t
\label{actionhhg}
\end{equation}%
where $a_{\mathrm{rec}}(\mathbf{k}+\mathbf{A}(t))=\left\langle
\mathbf{k}+\mathbf{A}(t)\right\vert \mathbf{d}\cdot \mathbf{e}_{x}\left\vert
\Psi \right\rangle $ and $a_{\mathrm{ion}}(\mathbf{k}+\mathbf{A}(t^{\prime
}))=\left\langle \mathbf{k}+\mathbf{A}(t^{\prime })\right\vert \mathbf{E(}%
t^{\prime }\mathbf{)\cdot r}\left\vert \Psi \right\rangle,$ are the recombination and ionization prefactors respectively. The terms $\mathbf{d}$, $\mathbf{e}_{x}$ and $\Omega $ denote  the dipole operator and the laser-polarization
vector and the
harmonic frequency, respectively. The ionization potential of the highest occupied molecular orbital, denoted by $\left\vert\Psi\right\rangle$, is given by $E_0$.  The length gauge is used throughout as the interference minima in the harmonic spectra vanish in the velocity gauge \cite{gaugepapers}, and the length form of the dipole operator is used as it accounts for the presence of the static dipole moment \footnote{Recently, we have shown that if a Gaussian basis set is employed in the modeling of the molecular orbitals, the velocity form of the dipole operator leads to a vanishing static dipole moment in heteronuclear molecule. This problem is absent if the length form is taken. For details see \cite{BradCarla}.}.

To solve Eq.~(\ref{amplhhg}) we use the saddle-point approximation \cite{atiuni}, where the action is stationary, such that $\partial S(t,t^{\prime },\Omega ,\mathbf{k}%
)/\partial t^{\prime }=\partial S(t,t^{\prime },\Omega
,\mathbf{k})/\partial t=0$ and $\partial S(t,t^{\prime },\Omega
,\mathbf{k})/\partial \mathbf{k=0} $.  This results in the
saddle-point equations

\begin{equation}
\frac{\left[ \mathbf{k}+\mathbf{A}(t^{\prime })\right] ^{2}}{2}+E
_{0}=0,  \label{tunnelsame}
\end{equation}%
\begin{equation}
\int_{t^{\prime }}^{t}d\tau \left[ \mathbf{k}+\mathbf{A}(\tau )\right] =%
\mathbf{0}.  \label{return}
\end{equation}%
and
\begin{equation}
\frac{\left[ \mathbf{k}+\mathbf{A}(t)\right] ^{2}}{2}+E _{0}=\Omega .
\label{recsame}
\end{equation}%

These equations have the following physical interpretation. Initially the electron is tunnel ionized from the binding potential. This results in Eq. (\ref{tunnelsame}), which gives the conservation of energy. Note that this equation has only complex solutions, as a consequence of the fact that tunneling has no classical counterpart.  Secondly, Eq. (\ref{return}) results in the electron momentum in the continuum being constrained such that the electron returns to its place of origin, which we take to be the geometric center of the molecule.   Finally, Eq. (\ref{recsame}) represents conservation of energy at recombination.
}

\subsection{Molecular orbitals and dipole matrix elements}
\label{molecular}
{\ The molecular orbitals can be calculated using the Linear Combination of
Atomic Orbitals (LCAO) approximation, along with the Born Oppenheimer
approximation. Under these assumptions, the molecular orbital wavefunction is given by
\begin{eqnarray}
\Psi(\mathbf{r})&=&\sum_{\alpha }c_{\alpha}^{(L)}\Phi
_{\alpha}^{(L)}(\mathbf{r}+\mathbf{R}/2)\nonumber\\
&+&(-1)^{l_{\alpha}-m_{\alpha
}+\lambda _{\alpha }}c_{\alpha}^{(R)}\Phi _{\alpha}^{(R)}(\mathbf{r}-%
\mathbf{R}/2)  \label{LCAOposition}
\end{eqnarray}
where \textbf{R} is the internuclear separation, $l_{\alpha }$ is the orbital quantum number and $ m_{\alpha }$ is the magnetic quantum number. We assume that the ions are positioned along the z axis, i.e., that $\mathbf{R}=R\hat{e}_z$. The
coefficients $c_{\alpha}^{(\xi )}$ form the linear
combination of atomic orbitals and are extracted from the quantum chemistry code GAMESS-UK \cite%
{GAMESS}. The indices $(L)$ and $(R)$ refer to the left or to the right
ion, respectively, which is a distinction required for heteronuclear molecules, whereas for homonuclear molecules, $c_{\alpha }^{(L)}=c_{\alpha
}^{(R)}=c_{\alpha}$ and $\Phi _{\alpha}^{(R)}=\Phi _{\alpha}^{(L)}=\Phi _{\alpha}$. The internuclear axis is taken to be along the $z$
direction, and the laser-field polarization is chosen along the radial
coordinate. The parameter $\lambda _{\alpha}$ determines the orbital symmetry of the molecular orbital, with $\lambda
_{\alpha }$=$|m_{\alpha}|$ and $\lambda _{\alpha}$=$|m_{\alpha}|$+1 for homonuclear molecules of
gerade and ungerade symmetry, respectively. }

The wavefunctions themselves are then expanded as Gaussian type orbitals
with a real 6-31 basis set, with
\begin{equation}  \label{contraction}
\Phi _{\alpha}^{(\xi )}(\mathbf{r})=\sum_{\nu }b_{\nu}^{(\xi
)}(r_{\chi })^{l_{\alpha }}e^{-\zeta _{\nu}^{(\xi )}r^{2}}.
\end{equation}%
We find that polarized basis sets make little difference to the orbital wavefunctions.
For the $\sigma ,$ $\pi _{x}$ and $\pi _{y}$ orbitals, $r_{\chi }=z,x$ and $y $, respectively. The contraction coefficients, $b_{\nu}^{(\xi )}$ and exponents, $\zeta _{\nu}^{(\xi )}$, are also obtained from GAMESS-UK \cite{GAMESS}, while the index $\xi $ distinguishes the left or right ion. Note that these
coefficients, along with the coefficients making up the LCAO, are real.

When using the strong-field approximation, all structural
information about the molecular orbital is contained in the recombination prefactor, $a_{\mathrm{rec}}(\mathbf{k}+\mathbf{A}(t))$, which is given by
\begin{equation}
a_{\mathrm{rec}}(\mathbf{k})=\frac{1}{(2\pi )^{3/2}}\int d^{3}r\mathbf{r}%
\cdot \hat{e}_{z}\exp [-i\mathbf{k}\cdot \mathbf{r}]\Psi (\mathbf{r}).
\end{equation}%
This is the component of $i\partial _{\mathbf{k}}\Psi (\mathbf{k})$ along the
laser-field polarization.  This prefactor gives rise to any two center interference which occurs in the harmonic spectra.  The ionization prefactor, $a_{\mathrm{ion}}(\mathbf{k}+\mathbf{A}(t^{\prime
}))$, on the other hand,  will mainly determine whether tunnel ionization will be suppressed or enhanced due to the shape of
a particular orbital.  Therefore it directly affects the overall harmonic yield, but is not responsible for any interference patterns.  Both prefactors are vanishing for nodes in the molecular orbitals.

In this work we use the momentum space wavefunctions to calculate the prefactors, which are given by%
\begin{eqnarray}
\psi(\mathbf{k})&=&\sum_{\alpha }\exp [i\mathbf{k\cdot }\frac{\mathbf{R}}{%
2}]c_{\alpha}^{(L)}\Phi _{\alpha}^{(L)}(\mathbf{k})\\
&+&(-1)^{l_{\alpha
}-m_{\alpha }+\lambda _{\alpha }}\exp [-i\mathbf{k\cdot }\frac{\mathbf{R}}{2}%
]c_{\alpha}^{(R)}\Phi _{\alpha}^{(R)}(\mathbf{k})\nonumber,
\label{orbitalpspace}
\end{eqnarray}%
where %
\begin{equation}
\Phi _{\alpha}^{(\xi )}(\mathbf{k})=\sum_{\nu }b_{\nu}^{(\xi )}\tilde{%
\varphi}_{\nu}^{(\xi )}(\mathbf{k}),
\end{equation}%
with
\begin{equation}
\tilde{\varphi}_{\nu}^{(\xi )}(\mathbf{k})=(-ik_{\beta })^{l_{\alpha }}%
\frac{\pi ^{3/2}}{2^{l_{\alpha }}\left( \zeta _{\nu}^{(\xi )}\right)
^{3/2+l_{\alpha }}}\exp [-k^{2}/(4\zeta _{\nu}^{(\xi )})].
\label{momentumwf}
\end{equation}%

Similarly to the position space expressions, $\beta =z$, $\beta =x$
and $\beta =y$ for the $\sigma $, $\pi _{x}$ and $\pi _{y}$
orbitals. The return condition (\ref{return}) guarantees that the
momentum $\mathbf{k}$ and the external field are collinear. Hence,
for a linearly polarized field the angle between the intermediate
momentum $\mathbf{k}$ and the internuclear axis $\mathbf{R}$ is
equal to the alignment angle $\theta _{L}$ between the molecule and
the field. The above-stated equations have been derived under the
assumption that only $s$ and $p$ states will be employed in order to
build the wavefunctions utilized in this work. For more general
expressions see Ref. \cite{CarlaBrad}.

\subsection{Interference Condition}
\label{interf}
{
Maxima and minima in the momentum-space wavefunction may be determined by
writing the exponents in Eq.~(\ref{orbitalpspace}) in terms of trigonometric
functions. In this case, one obtains%
\begin{equation}
\psi(\mathbf{k})=\sum_{\alpha }\mathcal{C}_{+}^{(\alpha )}\cos \left[
\frac{\mathbf{k\cdot R}}{2}\right] +i\mathcal{C}_{-}^{(\alpha )}\sin \left[
\frac{\mathbf{k\cdot R}}{2}\right] ,
\end{equation}%
with%
\begin{equation}
\mathcal{C}_{\pm }^{(\alpha )}=(-1)^{l_{\alpha }-m_{\alpha }+\lambda
_{\alpha }}c_{\alpha}^{(R)}\Phi _{\alpha}^{(R)}(\mathbf{k})\pm
c_{\alpha}^{(L)}\Phi _{\alpha}^{(L)}(\mathbf{k}).  \label{coeffinterf}
\end{equation}%
Calling $\vartheta =\arctan [i\mathcal{C}_{+}^{(\alpha )}/\mathcal{C}%
_{-}^{(\alpha )}],$we find
\begin{equation}
\psi(\mathbf{k})=\sum_{\alpha }\sqrt{\left( \mathcal{C}_{+}^{(\alpha
)}\right) ^{2}-\left( \mathcal{C}_{-}^{(\alpha )}\right) ^{2}}\sin
[\vartheta +\mathbf{k\cdot R}/2].  \label{pspaceinterf}
\end{equation}%
Eq.~(\ref{pspaceinterf}) exhibit minima for $\vartheta +\mathbf{k\cdot R}%
/2=n\pi$.

Note, however, that the coefficients $\mathcal{C}_{\pm }^{(\alpha )}$
defined in Eq.~(\ref{coeffinterf}) depend on the wavefunctions at the left
and right ions. Since these wavefunctions themselves depend on the momentum $%
\mathbf{k}$, one expects the two-center patterns to be blurred for
heteronuclear molecules. In contrast, for homonuclear molecules,
$c_{\alpha}^{(L)}=c_{\alpha}^{(R)}$ and $\Phi
_{\alpha}^{(L)}(\mathbf{k})=\Phi _{\alpha}^{(R)}(\mathbf{k})$. This
implies that the momentum dependence in the argument $\vartheta$
cancels out and that the interference condition in
Refs.~\cite{Milosevic,CarlaBrad} is recovered. In this case, sharp
interference fringes are expected to be present.

The above-stated condition does not only lead to interference fringes in the
bound-state momentum wavefunctions but also in the high-harmonic spectra.
This is due to the fact that the dipole matrix elements depend on the
wavefunctions (\ref{orbitalpspace}). In this latter case, $\mathbf{%
k\rightarrow k+A}(t)$ in Eqs.~(\ref{orbitalpspace})-(\ref{pspaceinterf}).

For a linearly polarized monochromatic field of frequency $\omega$, using the saddle-point
equation (\ref{recsame}), the generalized interference condition (\ref%
{pspaceinterf}) may be expressed in terms of the harmonic order $n$ as%
\begin{equation}
n=\frac{E_{0}}{\omega }+\frac{2(\kappa\pi -\vartheta )^{2}}{\omega R^{2}\cos
^{2}\theta _{L}},  \label{interfomega}
\end{equation}%
where $E_{0}$ is the absolute value of the bound-state energy in question, $%
\kappa $ is an integer number, $\theta _{L}$ is the alignment angle, $R$ is
the internuclear distance and $\vartheta $ is defined above). This
interference condition has been first derived in \cite{DM2009} and has been used by us in previous work \cite{CarlaBrad}.

  In the specific case of orbitals of $\pi_{g}$ type, which are constructed entirely from p-type atomic orbitals such that $l_{\alpha}=1$, are of gerade symmetry such that $\lambda_{\alpha}=0$, and are made from a homonuclear species such that $\Phi _{\alpha}^{(R)}(\mathbf{k})=\Phi _{\alpha}^{(L)}(\mathbf{k})$ and $c_{\alpha}^{(R)}=c_{\alpha}^{(L)}$, one finds that $\vartheta=0$.  This implies that for the zeroth order interference, where $\kappa=0$, the second term of (\ref{interfomega}) will vanish leaving the angular independent term
 \begin{equation}
n=\frac{E_{0}}{\omega }.
 \label{interfomega1}
\end{equation}%
This minimum, however, would occur exactly at the ionization threshold, for which the present framework is expected to break down. Hence, it will not be discussed here.}
\section{Results}

\label{results}

\subsection{Position and momentum space wavefunctions}

\label{wfunctions}

In this section, we will display the specific molecular orbitals
employed by us in the results that follow. We choose two pairs of molecules
composed of a homonuclear molecule and an isoelectronic heteronuclear
molecule. The first pair, {$\mathrm{Be}_{2}$ and LiB, possess eight electrons and
a }$\sigma $ highest occupied molecular orbital, while the second pair,
$\mathrm{O}_{2}$ and NF, have sixteen electrons and a $\pi $ highest occupied
molecular orbital.

\begin{figure}[tbp]
\begin{center}
\includegraphics[width=8cm]{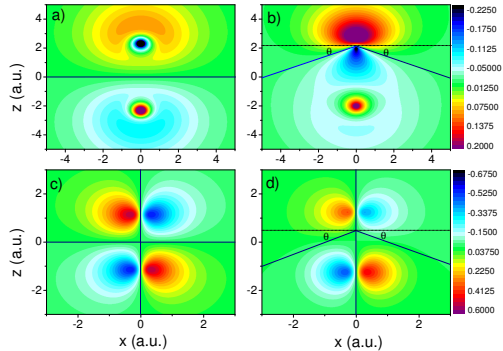}
\end{center}
\caption{Highest occupied molecular orbitals in position space for
the homonuclear molecules $\mathrm{Be}_{2}$ and $\mathrm{O}_{2}$
[panels a) and c)], and their heteronuclear isoelectronic
counterparts LiB and NF [panels b) and d)].  The orbital symmetries
in panels a), b), c) and d) are $2\sigma_u$, $4\sigma$, $1\pi_g$ and
$2\pi$, respectively, and the bond lengths are
$R^{(\mathrm{Be}_{2})}=4.642$ a.u.,  $R^{(\mathrm{LiB})}=4.642$
a.u., $R^{(\mathrm{O}_2)}=2.280$ a.u., and $R^{(\mathrm{NF})}=2.485$
a.u. } \label{position}
\end{figure}

{In Fig. \ref{position}, we exhibit the HOMO for homonuclear molecules, $%
\mathrm{Be}_{2}$ and $\mathrm{O}_{2}$, and heteronuclear molecules,
LiB and NF, as obtained with GAMESS-UK
[Figs.~\ref{position}.(a),(b),(c) and (d) respectively]. The
internuclear axis is aligned along the $z$-axis. The position-space
wavefunctions of $\mathrm{Be}_{2}$ and LiB are of $\sigma $ type
with $\mathrm{Be}_{2}$ having ungerade symmetry, $\sigma _{u}$, and
a nodal plane along the $x$-axis at $z=0$. For LiB, there is a bias
towards the Boron atom. One can also clearly see the contribution of
the p type atomic orbital
which is introduced by the Boron atom. The position-space wavefunctions of $%
\mathrm{O}_{2}$ and NF are of $\pi $ type with $\mathrm{O}_{2}$
having gerade symmetry, $\pi _{g}$, and a nodal plane along the
x-axis at $z=0$ and the z-axis at $x=0$. NF shows a bias towards the
Fluorine atom. }

\begin{figure}[tbp]
\begin{center}
\includegraphics[width=9cm]{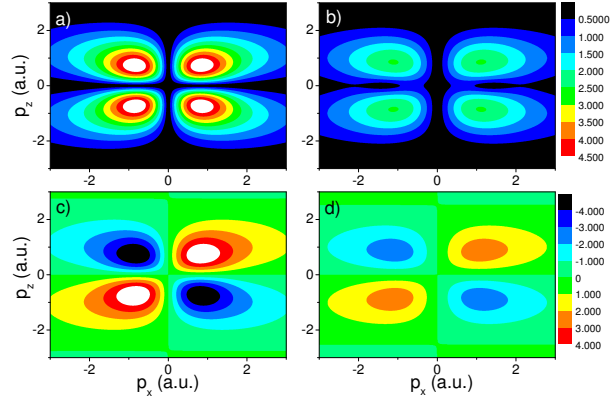}
\end{center}
\caption{Absolute values and real parts of the highest occupied molecular
orbitals in momentum space for homonuclear molecule $\mathrm{O}_{2}$ [panels a) and
c) respectively] and its heteronuclear counterpart NF [panels b) and d)]. }
\label{momentum2}
\end{figure}

\begin{figure}[tbp]
\begin{center}
\includegraphics[width=9cm]{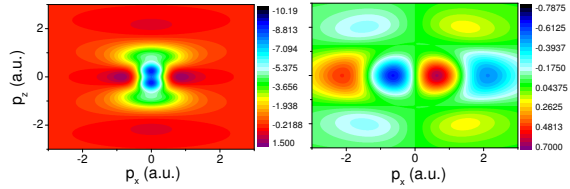}
\end{center}
\caption{Real part of the highest occupied molecular orbital in momentum
space for LiB and imaginary part of the highest occupied molecular orbital
in momentum space for NF, panels a) and b) respectively.}
\label{momentum3}
\end{figure}

We have found that both momentum-space wavefunctions, given in Figs. \ref%
{momentum1}, \ref{momentum2} and \ref{momentum3}, are symmetric with
respect to $(p_{x},p_{z})\rightarrow (-p_{x},-p_{z})$ in comparison
to their position-space counterparts. Despite that, the
momentum-space wavefunctions for homonuclear and heteronuclear
molecules exhibit different properties, especially if a real basis
set is used in the construction of the position-space wavefunctions
$\psi(\mathbf{r})$. They come from the properties of Fourier
transforms and will determine whether a particular node will be
blurred in the heteronuclear case. An even wavefunction in position
space will lead to a real wavefunction in momentum space, as the
Fourier transform of a real and even function should be real. For
the same reason, if the position-space wavefunction is odd, the
corresponding momentum-space wavefunction should be pure imaginary.
Neither property will hold if the position-space wavefunction does
not possess a well-defined symmetry. In this case, its
momentum-space counterpart will be complex.

\begin{figure}[tbp]
\begin{center}
\includegraphics[width=9cm]{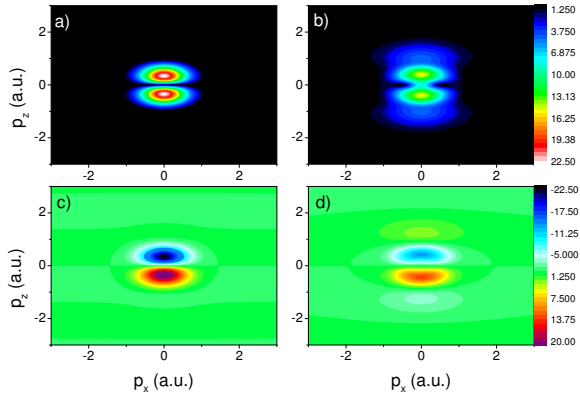}
\end{center}
\caption{Absolute values and imaginary parts of the highest occupied
molecular orbitals in momentum space for the homonuclear molecule $\mathrm{Be}_{2}$
[panels a) and c) respectively] and its heteronuclear counterpart LiB
[panels b) and d)]. }
\label{momentum1}
\end{figure}

We will now have a closer look at the nodes in the momentum-space
wavefunctions. For $\mathrm{Be}_{2}$, the momentum-space
wavefunction of the $\sigma _{u}$ HOMO is pure imaginary, and
exhibit a clear node along the $p_{x}$-axis (see
Figs.~\ref{momentum1}.(a) and (c) for $|\psi(\mathbf{k})|$ and
$\mathrm{Im}[\psi(\mathbf{k}]$, respectively). This node is caused
by the linear combination of atomic orbitals giving rise to the
$\sigma _{u}$ orbital.

The above-mentioned node is lost in the absolute value of the HOMO
wavefunction of LiB, shown in Fig.~\ref{momentum1}.(b), even though
a suppression along the $p_{x}$ axis is still present.  Apart from
that, there is a secondary set of peaks extending towards higher
momentum values. One should note, however, that a very clear central
node for LiB at the $p_{x}$ axis can be seen in the imaginary part
of corresponding wavefunction, shown in Fig.~\ref{momentum1}.(d).
This feature is blurred by the real part of $\psi(\mathbf{k})$,
which does not have this node, and is of comparable magnitude (see
Fig.~\ref{momentum3}.(a)).  As discussed above, a complex
wavefunction in momentum space is related to an asymmetric
wavefunction in position space. Hence, the loss of symmetry in
position space is what leads to the loss of the clear nodal plane in
momentum space. Physically, the asymmetry in the position-space
wavefunction can be traced back to the static dipole moment along
the internuclear axis.

In comparing $\mathrm{O}_{2}$ and NF momentum space wavefunctions as shown in Fig. %
\ref{momentum2}, one can see that the features are quite similar for one of
the nodal planes. Specifically, the well defined nodal plane along the $%
p_{x}$-axis observed for $\mathrm{O}_{2}$ has been slightly blurred
for NF (see Figs.~\ref{momentum2}.(a) and (b), respectively). This
again is because of the loss of symmetry in the heteronuclear case,
due to the static dipole moment. Note once more that an even and
real position-space wavefunction, such as the HOMO in
$\mathrm{O}_{2},$ leads to a real momentum-space wavefunction. This
is not the case for NF, and the non-vanishing imaginary part of the
momentum space wavefunction gives rise to the blurring. However, it
can be seen from Fig.~\ref{momentum3}.(b) that the imaginary part of
the momentum space wavefunction for NF is much smaller than its real
part. This implies that the NF molecule is quite close to having
gerade type symmetry, and therefore one would expect similar
features in the position and momentum space wavefunctions as
compared to $\mathrm{O}_{2}$.

In contrast, the other nodal plane, along the $z$-axis, remains completely
unaffected by the presence of the static dipole moment. Indeed, this nodal
plane is just as pronounced in both molecules, irrespective of being
homonuclear or heteronuclear. Mathematically, this is because one can define
two types of nodal planes in position space: one of which is due to the
characteristic of the $p$-type atomic orbitals, which are of the form $%
\sum_{\nu }b_{\nu}^{(\xi )}xe^{-\zeta _{\nu}^{(\xi )}r^{2}}$, and
therefore disappear at $x=0$, irrespective of the contraction
coefficients and exponential coefficients, and the other of which is
due to the linear combination of atomic orbitals. The latter results
in a cancelation for a homonuclear molecule, and hence a nodal
plane. This however does not occur for a heteronuclear molecule. The
nodal plane along the $p_{z}$-axis is of the first type, and is
therefore present for both molecules, and the nodal plane along the
$p_{x}$-axis is of the second type and not present for NF. This
argument can be mapped into the momentum space using Fourier
transforms (see Eq.~(\ref{orbitalpspace})).

\subsection{Harmonic Spectra}
\label{spectra}

\begin{figure}[tbp]
\begin{center}
\includegraphics[width=9cm]{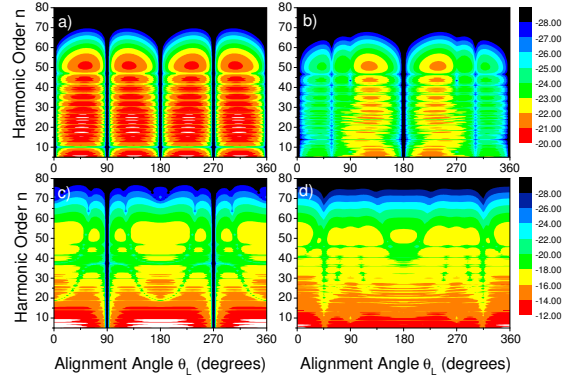}
\end{center} \caption{Harmonic spectra as a function of alignment angle for
the molecule a) $\mathrm{Be}_{2}$ of ionization potential 0.2390
a.u., b) $LiB$ of ionization potential 0.1942 a.u., c) $O_{2}$ of
ionization potential 0.2446 a.u. and d) NF of ionization potential
0.2246 a.u. A linearly polarized laser field of frequency
$\protect\omega$=0.057 a.u. and intensity $I=4 \times
10^{14}\mathrm{W/cm}^2$ is used. The spectra have been computed
using the length gauge and employing the dipole operator in the
length form.} \label{contour1}
\end{figure}
We now present the harmonic spectra computed in  a linearly polarized monochromatic wave of frequency $\omega$ and amplitude $\omega$$A_{0}$, for all molecules discussed previously. These results are displayed in Fig. \ref{contour1}. It can be seen that all spectra contains a series of
cuts, i.e., a complete suppression of the yield for particular angles. Most
notably in the case of $\mathrm{O}_{2}$ these occur at alignment angles $\theta _{L}$%
=0,90,180,270 and 360 degrees (see Fig.~\ref{contour1}.(a)). By
observing the position space wavefunction in Fig.~\ref{position}.(c)
and the absolute value of the momentum space wavefunction in
Fig.~\ref{momentum2}.(a) this can immediately be attributed to the
nodal planes of the molecular wavefunction, which occur at these
angles, and therefore prohibit ionization or recombination. The
situation in the case of NF , however, is modified, as shown in
Fig.~\ref{contour1}.(b). The harmonic signal completely vanishes at
alignment angles $\theta _{L}$=0,180 and 360 degrees, because at
these angles the nodal planes in the highest occupied molecular
orbital occur due to the nature of the $p$-type atomic orbitals.
However, due to the polar nature of these heteronuclear molecules,
and a linear combination of atomic orbitals which no longer cancels
out, the nodal planes at alignment angles $\theta _{L}$=90 and 270
degrees are no longer present.

\begin{figure}[tbp]
\begin{center}
\includegraphics[width=9cm]{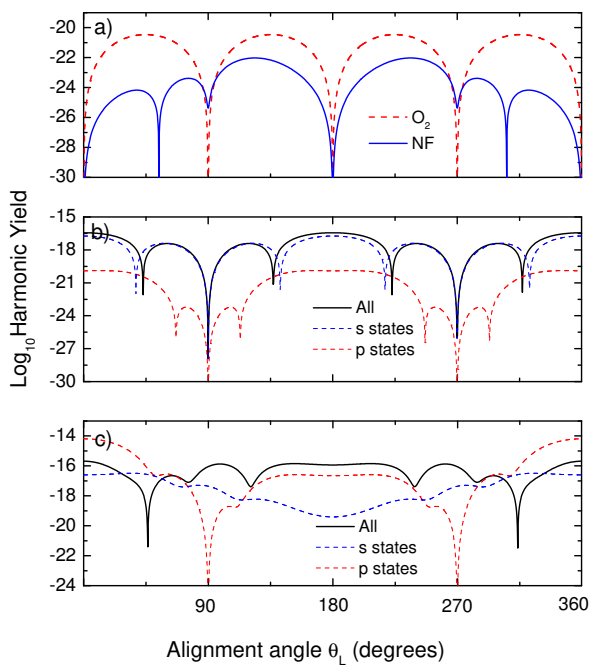}
\end{center}
\caption{The harmonic yield of a) $\mathrm{O}_{2}$ and NF for the
$45^{th}$ harmonic
vs the alignment angle, b) s and p orbitals to the harmonic yield of $%
\mathrm{Be}_{2}$ for the $25^{th}$ harmonic vs the alignment angle
and c) s and p orbitals to the harmonic yield of LiB for the
$25^{th}$ harmonic vs the alignment angle.  All parameters are the
same as in the previous figure.} \label{harmoniccuts}
\end{figure}

In fact, several differences with regard to the homonuclear case are observed. Firstly, the signal is no longer completely
suppressed, but there is a distinct minimum which occurs at alignment angles $%
\theta _{L}$= 54 and 306 degrees. There is also a second minimum, but not a
cut, beyond harmonic n=35, which occurs at the alignment angles $\theta _{L}$%
=90 and 270 degrees, that is, where the nodes were in the
homonuclear molecule. We attribute this to the remnants of the nodal
plane in its heteronuclear counterpart. This can be seen more
clearly in Fig.~\ref{harmoniccuts}.(a), where the harmonic spectra
of $\mathrm{O}_2$ and NF are plotted as functions of the alignment
angle for fixed harmonic order.

Inspection of the position and momentum space wavefunction suggests
that the polar nature of the heteronuclear molecules deforms the
nodal plane such that, although there is no longer a node, there is
a suppression which occurs at the angle to which the plane has been
deformed. As the molecule is rotated the new minimum in the
wavefunction is first experienced before $\theta_{L}$=90 degrees but
then occurs at the same angle after $\theta_{L}$=270 degrees.
Therefore the spectra has reflectional symmetry about an alignment
angle of 180 degrees. There is no longer a clear nodal plane, as the
probability density associated to the wavefunction is small, but
nonvanishing. Hence, there is not a complete suppression in the
harmonic spectra. However, a clear minimum is still visible.

Note that, for $\mathrm{O}_{2}$, the two-center interference maxima
and minima predicted by Eq. (\ref{interfomega1}) are located either
beyond the cutoff or at the ionization threshold. Therefore, they do
not occur in the parameter range of interest. This also holds for
its heteronuclear counterpart.

Now observing the spectra for $\mathrm{Be}_{2}$, displayed in Fig. \ref{contour1}.(c), one may identify nodes at $%
\theta_{L}$=90 and 270 degrees, as one would expect by observing the
position and momentum space wavefunctions. In the spectra of the
heteronuclear counterpart LiB, presented in Fig. \ref{contour1}.(d), these nodes have been replaced by minima at $%
\theta_{L}$=45 and 315 degrees, for the same reasons as described above.

The spectra for $\mathrm{Be}_{2}$ also exhibit two-center
interference minima, corresponding to $\kappa=0$, $\kappa=1$ and
$\kappa=2$ in Eq.~(\ref{interfomega}), whose energy position depends
strongly on the alignment angle. These are shifted
considerably in the heteronuclear counterpart LiB. This is because the $%
\mathrm{Be}_{2}$ molecular wavefunction is constructed almost
entirely of $s$-type atomic orbitals, whereas in the case of LiB the
Boron atom introduces $p$-type atomic orbitals into the molecular
wavefunction and hence changes the interference condition. Note
that, with regard to $\mathrm{Be}_2$, there is a blurring in the
two-center patterns for LiB. This is expected according to the
discussion in Sec.~\ref{interf}.
\begin{figure}[tbp]
\begin{center}
\includegraphics[width=9cm]{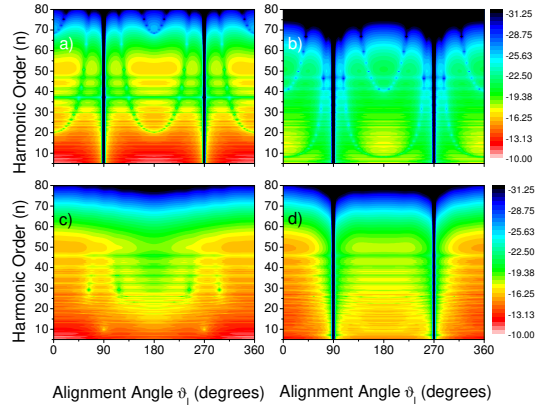}
\end{center}
\caption{Harmonic spectra as a function of alignment angle for a)
the s-type orbitals of the $\mathrm{Be}_{2}$ HOMO, b) the p-type
orbitals of the $\mathrm{Be}_{2}$ HOMO, c) the s-type orbitals of
the $LiB$ HOMO, d) the p-type orbitals of the LiB HOMO, for the same
parameters as in the previous figure.} \label{contour2}
\end{figure}

This is shown in more detail in Fig.~\ref{contour2}, in which the
individual contributions of the $s$ and $p$ states to the spectra of
$\mathrm{Be}_{2}$ and LiB are presented (upper and lower panels,
respectively). For these individual contributions, there is also a
considerably simpler interference condition for homonuclear
molecules, which can be obtained from Eq. (\ref{interfomega}) by
setting either $C_{+}$ or $C_-$ equal to zero. Specifically the
$s$-state and $p$-state contributions for an ungerade orbital
exhibit minima at the harmonic frequencies
\begin{equation}
\Omega^{(u)}_s=E_0+\frac{2\kappa^2\pi^2}{R^2\cos^2\theta_L}
\label{interfsungerade}
\end{equation}
and
\begin{equation}
\Omega^{(u)}_p=E_0+\frac{(2\kappa+1)^2\pi^2}{2 R^2\cos^2\theta_L},
\label{interfpungerade}
\end{equation}
respectively, where $\kappa$ is a integer. In
Fig.~\ref{contour2}.(a), we may identify the minima for $\kappa=1$
and $\kappa=2$ according to Eq. (\ref{interfsungerade}), while in
Fig.~\ref{contour2}.(b) the minima corresponding to $\kappa=0$ and
$\kappa=1$ can be easily found. If the s p mixing is considered, the
full expression (\ref{interfomega}) must be used.

The contributions from $p$-type orbitals to the harmonic spectrum of
$\mathrm{Be}_{2}$ are orders of magnitude smaller than the
contribution of p-type orbitals in LiB. Therefore the overall
spectrum of $\mathrm{Be}_{2}$ is dominated mostly by the s-type
atomic orbitals. In contrast, in LiB, both types of orbitals lead to
comparable contributions to the spectra, and the $s$ $p$ mixing will
influence the energy position of the overall two-center patterns.
Apart from that, the ionization potentials and internuclear
separation of both molecules are slightly different. This will
influence the two-center interference further.

The effects of $s$ $p$ mixing are also displayed in Fig.
\ref{harmoniccuts}.(b) and Fig.~\ref{harmoniccuts}.(c) for a fixed
mid-plateau harmonic. The figure clearly shows that where one
observes that, for $\mathrm{Be}_2$, the $s$-type orbitals dominate
and the main effect of the $p$-type orbitals is to introduce a small
shift in the interference minima. On the other hand, for LiB, the
overall maxima and minima are considerably altered by $s$ $p$
mixing.  In the figure, one can also see that the nodal planes
occurring at alignment angles $\theta_{L}$=90 and 270 degrees in
$\mathrm{Be}_{2}$, are shifted by the contribution of the s-type
orbitals in LiB.

Finally, we will comment on the overall harmonic intensities.
 The harmonics generated from NF are less intense than those from $\mathrm{O}_{2}$. As
the ionization potential of NF is slightly lower than that of
$\mathrm{O}_{2}$, one would expect more ionization and therefore
higher harmonic yield. This implies that the shape of the molecular
wavefunction results in suppression at the recombination step for
NF, an effect which is more prominent than the difference in
ionization potential. Harmonics generated from $\mathrm{Be}_{2}$ and
LiB are much more intense than those generated by $\mathrm{O}_{2}$
and NF, despite similar ionization potentials. This suggests that
the presence of nodal planes, or minima associated with strong
suppression of the wave function, reduce the yield of the entire
spectrum. As $\mathrm{O}_{2}$ has two nodal planes and
$\mathrm{Be}_{2}$ has one, harmonic spectra from the former is less
intense than that of the latter.
\section{Conclusions}
\label{conclusions} In this work, we investigated the dependence of
the high-order harmonic spectra on the alignment angle between the
diatomic molecules and the laser-field polarization, for
isoelectronic pairs consisting of a homonuclear and a heteronuclear
molecule. We employed a single active electron approximation, using
the HOMO as the active orbital, within the strong-field
approximation.

Our studies lead us to conclude that there are two types of nodes in the bound-state orbitals of molecules, which will lead to a strong suppression in the harmonic yield when such nodes are parallel to the laser--field polarization. The first type is caused by the nodes in the atomic orbitals at \emph{each} ion, and is present at the same location for homonuclear molecules and their heteronuclear counterparts. The second type of node is due to the sum or subtraction of atomic orbitals at \emph{different} centers within the linear combination of atomic orbital (LCAO) approximation. Both types of minima are present for homonuclear molecules. For their heteronuclear counterparts, however, the asymmetry of the molecule eliminates the latter nodes. This implies that the imprints of the first type of nodes in the harmonic spectra are common to isoelectronic homonuclear and heteronuclear molecules, and could in principle be observed in both cases.

In general, the asymmetry also leads to some blurring in the interference patterns caused by high-harmonic emission at spatially separated centers in the molecule. Furthermore, depending on the molecule, $s$ $p$ mixing will be different. This will lead to shifts in the energy positions of the two-center patterns. On a more technical level, we have also been able to map the symmetry or asymmetry of the molecular orbitals in position space to properties of their momentum-space counterparts.

The above-stated findings show that nodal planes in a heteronuclear
molecule can be related to those in an isoelectronic homonuclear
molecule. Hence, in principle, the latter can be used as a reference
point in order to understand the behavior of the former, by using
high-order harmonic generation. The shifts in the nodal planes due
to the distortions in the wavefunctions or the absence thereof can
be mapped into features in the high-harmonic spectra. This may shed
some light in the imaging of heteronuclear molecules.

\ack We thank H. J. J. van Dam, P. J. Durham, P. Sherwood and J.
Tennyson for very useful discussions. This work was supported by the
UK Engineering and Physical Sciences Research Council (EPSRC), and
by the Daresbury Laboratory.

\end{document}